\documentclass[aps,prl,twocolumn,groupedaddress,showpacs]{revtex4}
\usepackage{amsmath}  
\usepackage{amsfonts} 
\usepackage{amssymb}
\usepackage{graphicx} 

 \usepackage{lstdoc,longtable}
 \usepackage{tabularx}
 \usepackage{stix}

\begin{document}

\title{Relativistic relative velocities and relativistic acceleration}

\author{Grzegorz M. Koczan\footnote{grzegorz\_koczan@sggw.edu.pl, gkoczan@fuw.edu.pl}}

\affiliation{Warsaw University of Life Sciences (WULS/SGGW), Poland }

  

\date{\today}

\begin{abstract}
 It turns out that the standard application of the four-vector SR formalism does not include the concept of relative velocity. Only the absolute velocity is described by the four-vector, and even the Lorentz transformation parameters is described by the three-dimensional velocity.
 This gap in the development of the SR formalism reflects the lack of some significant velocity subtraction operations. The differential application of these operations leads to a relativistic acceleration.

\


\end{abstract}








%




\maketitle 



\section{1. Introduction}
Relative velocity is the defined difference of two velocities. In the theory of relativity, such a difference is most easily realized using the Lorentz transformation
 \cite{Rindler}:
\begin{equation}
\label{Lorentz}
x'=\gamma_{\mathrm{v}} (x-\mathrm{v}t) \:,\:t'=\gamma_{\mathrm{v}} \left(t - \frac{\mathrm{v}}{c^{2}}\,x \right) \:,\: y'=y \:,\: z'=z,
\end{equation}
where: $\gamma_{\mathrm{v}}=1/\sqrt{1-\mathrm{v}^2/c^2}$.
The Lorentz transformation of velocity is some way of subtracting velocities:
\begin{equation}
\label{velocity x}
\mathrm{u}_x'=\frac{dx'}{dt'}=\frac{\mathrm{u}_x-\mathrm{v}}{1-\mathrm{u}_x\mathrm{v}/c^2}=:\mathrm{u}_x\ominus \mathrm{v},
\end{equation}
\begin{equation}
\label{velocity y}
\mathrm{u}_y'=\frac{dy'}{dt'}=\mathrm{u}_y\frac{\sqrt{1-\mathrm{v}^2/c^2}}{1-\mathrm{u}_x\mathrm{v}/c^2}=:\mathrm{u}_y\boxminus_{\mathrm{u}_x} \mathrm{v}.
\end{equation}
The component parallel to the boost velocity is subject to the $\ominus$ subtraction, and the perpendicular components are subject to the $\boxminus_{\mathrm{u}_x}$ operation, which is parameter dependent and is more like multiplication than subtraction.
If we apply appropriate operations for all components at the same time $(\ominus,\boxminus_{\mathrm{u}_x},\boxminus_{\mathrm{u}_x})$ then we will get the full vector subtraction $\ominus_0$ of Einstein velocity \cite{Einstein}:
\begin{equation}
\label{velocity Einstein}
\mathbf{u'}=\frac{\mathbf{u}-\mathbf{v}+\frac{\gamma_{\mathrm{v}}}{\gamma_{\mathrm{v}}+1}(\mathbf{u}\times\mathbf{v})\times\mathbf{v}/c^{2}}{1-\mathbf{u}\mathbf{v}/c^{2}}=:\mathbf{u}\ominus_0\mathbf{v}.
\end{equation}
Originally, Einstein, like many behind him (but not all \cite{Cannoni, Fock}), considered addition, not subtraction. However, the addition here is less natural and leads to an unequally ambiguous sequence
  $\mathbf{u}\oplus\mathbf{v}=\mathbf{u}\ominus_0\mathbf{(-v)}$ vs $\mathbf{u}\oplus\mathbf{v}=\mathbf{v}\ominus_0\mathbf{(-u)}$ or $\mathbf{v}\oplus\mathbf{u}=\mathbf{u}\ominus_0\mathbf{(-v)}$. The latter convention is most often used \cite{Mexico, Ungar}, which changes the right-hand character of the operation to the left-hand one. 

It is also worth calculating the velocity differential in relation to the velocity subject to the boost (compare \cite{Rebilas}):
\begin{equation}
\label{dif x}
d\mathrm{u}_x':=\frac{\partial \mathrm{u}_x'}{\partial \mathrm{u}_x}\Big|_{\mathrm{u}_x=\mathrm{v}}d\mathrm{u}_x=\frac{d\mathrm{u}_x}{1-\mathrm{v}^2/c^2}=\gamma^2_{\mathrm{v}}d\mathrm{u}_x,
\end{equation}
\begin{equation}
\label{dif y}
d\mathrm{u}_y':=\frac{\partial \mathrm{u}_y'}{\partial \mathrm{u}_x}\Big|_{\mathrm{u}_y=0}d\mathrm{u}_x+\frac{\partial \mathrm{u}_y'}{\partial \mathrm{u}_y}\Big|_{\mathrm{u}_x=\mathrm{v}}d\mathrm{u}_y=\gamma_{\mathrm{v}} d\mathrm{u}_y.
\end{equation}
In vector terms, this equation takes the form (see \cite{Dragan AJP}):
\begin{equation}
\label{du0}
d\mathbf{u}'
:=(d\mathbf{u}')_{\mathbf{v}}
:=\frac{\partial\mathbf{u}'}{\partial\mathrm{u}_i}\Big|_{\mathbf{u}=\mathbf{v}}d\mathrm{u}_i
=\gamma_{\mathrm{v}} d\mathbf{u}+\frac{\gamma^3_{\mathrm{v}}}{\gamma_{\mathrm{v}}+1}\frac{\mathbf{v}(\mathbf{v}d\mathbf{u})}{c^2}.
\end{equation}
It turns out that the differential calculated with respect to the second variable is basically only different in sign:
\begin{equation}
\label{dv0}
(d\mathbf{u}')_{\mathbf{u}}
:=\frac{\partial\mathbf{u}'}{\partial\mathrm{v}_i}\Big|_{\mathbf{v}=\mathbf{u}}d\mathrm{v}_i
=-\gamma_{\mathrm{u}} d\mathbf{v}-\frac{\gamma^3_{\mathrm{u}}}{\gamma_{\mathrm{u}}+1}\frac{\mathbf{u}(\mathbf{u}d\mathbf{v})}{c^2}.
\end{equation}

At high speeds, it is natural that the $d\mathrm{u}_x$ differential, which is also the differential of $\mathrm{u}$ value, scales as in (\ref{dif x}) with the appropriate power of the gamma factor \cite{Barrett}. Such scaling is necessary so that the transformed speed value does not exceed the speed of light. On the other hand, other scaling of the perpendicular speed differential (\ref{dif y}) results from time dilation between transformed systems. In \cite{Rebilas 3} an attempt was made to reconcile the scaling of both differentials. However, the scaling of the perpendicular differential does not follow the idea of an essentially directional (axial) change in the velocity vector - an analogous discrepancy is included in (\ref{velocity y}).

Another disadvantage of Einstein subtraction (\ref{velocity Einstein}) is the lack of its antisymmetry and the lack of associativity for the addition cooperation \cite{Ungar, Ahmad Alam, groupoid}.
As we will see, with the right approach to Lorentz boosts, the problem of the lack of velocity subtraction antisymmetry can be eliminated. 
In a sense, the problem of lack of antisymmetry also does not occur at the differential level (\ref{du0},\ref{dv0}). On the other hand, the problem of the lack of velocity addition associativity is more serious and is not covered by this paper. This problem has been attacked in the works \cite{Ahmad, groupoid, Kocik}. It should be emphasized that the lack of associativity explicitly concerns the velocity composition, not the Lorentz group. 

Although the four-vector approach is not considered in the SR for the velocity subtraction, let us consider such the four-velocities \cite{Minkowski, Wroblewski}:
\begin{equation}
\label{fourvectors}
    \upsilon^{\mu}=(\gamma_{\mathrm{u}} c, \gamma_{\mathrm{u}} \mathbf{u})\:,\:
    \nu^{\mu}=(\gamma_{\mathrm{v}} c, \gamma_{\mathrm{v}} \mathbf{v}).
\end{equation}
Let's try to provisionally write the equations (\ref{velocity x}, \ref{velocity y}) by subtracting four-velocities:
\begin{equation}
\label{naiwne}
    \frac{\upsilon^{\mu}-\nu^{\mu}}{\upsilon \circ \nu/c^2}=\frac{(\gamma_{\mathrm{u}}c-\gamma_{\mathrm{v}}c\:,\:\gamma_{\mathrm{u}} \mathbf{u}-\gamma_{\mathrm{v}} \mathbf{v})}
    {\gamma_{\mathrm{u}}\gamma_{\mathrm{v}}(1-\mathbf{uv}/c^2)},
\end{equation}
where: $\upsilon\circ\nu=\upsilon_{\beta}\nu^{\beta}$ is the scalar product of four-vectors.
Unfortunately, this formula is not valid for the parallel component (\ref{velocity x}), although it is correct for the perpendicular component
 (\ref{velocity y}). Nevertheless, it turns out that the equation (\ref{velocity Einstein}) can be formally converted into a four-vector form with the appropriate tools (see \cite{binary, Oziewicz}). However, this paper focuses on other (as it will turn out to be valid) methods of velocity subtraction that are ideologically similar to the equation (\ref{naiwne}).

\section{2. Axial subtraction of velocities}

As indicated above, the orthodox velocity subtraction method (\ref{velocity Einstein}) has several natural disadvantages.
The most artificial is the operation of $\boxminus_{\mathrm {u}_x} $ for components perpendicular to the subtrahend velocity, which differs significantly from the operation of $\ominus$ for parallel components.
Therefore, we can try to standardize the operation of $\ominus$ for all components \cite{Koczan 2019, Mexico}:
\begin{equation}
\label{axial x}
\mathrm{w}_{|x}:=\mathrm{u}_x\ominus \mathrm{v}=\frac{\mathrm{u}_x-\mathrm{v}}{1-\mathrm{u}_x\mathrm{v}/c^2},
\end{equation}
\begin{equation}
\label{axial yz}
\mathrm{w}_{|y}:=\mathrm{u}_y\ominus \mathrm{v}_y=\mathrm{u}_y\ominus 0=\mathrm{u}_y,
\end{equation}
\begin{equation}
\label{axial z}
\mathrm{w}_{|z}:=\mathrm{u}_z\ominus \mathrm{v}_z=\mathrm{u}_z\ominus 0=\mathrm{u}_z.
\end{equation}
If the $x$ axis goes along $\mathbf{v}$ then the above equations define an equivalent vector operation $\ominus_{\parallel}=(\ominus,\ominus,\ominus)$:
\begin{equation}
\label{axial}
\mathbf{u}\ominus_{\parallel}\mathbf{v}:=\frac{\mathbf{u}-\mathbf{v}+\frac{\mathbf{uv}}{\mathrm{v}^2}(\mathbf{u}\times\mathbf{v})\times\mathbf{v}/c^2}{1-\mathbf{u}\mathbf{v}/c^2}=:\mathbf{w}_|.
\end{equation}
This subtraction only changes the parallel component to the velocity subtrahend, so it will be called axial subtraction and $\mathbf{w}_|$ relative axial velocity.

STATEMENT 1

Axial subtraction of velocity vectors (\ref{axial}) is of the form:
\begin{equation}
\label{axial phi}
\mathbf{u}\ominus_{\parallel}\mathbf{v}=\mathbf{w}_|=\mathbf{u}-\varphi\mathbf{v},
\end{equation}
where the $\varphi$ function results from parallel subtraction (\ref{axial x}).

Proof. Using (\ref{axial x}) to (\ref{axial phi}) allows to calculate:
\begin{equation}
\label{phi}
\varphi=\frac{1-\mathrm{u}^2_x/c^2}{1-\mathrm{u}_x\mathrm{v}/c^2}=\frac{1-(\mathbf{uv})^2/(c\mathrm{v})^2}{1-\mathbf{u}\mathbf{v}/c^2},
\end{equation}
which inserted into (\ref{axial phi}) after simplifications leads to axial subtraction (\ref{axial}), Q.E.D.

Calculating the differentials of relative axial velocity is not difficult:
\begin{equation}
\label{dif axial}
d\mathrm{w}_{|x}=\gamma^2_{\mathrm{v}} d\mathrm{u}_x, \:\:d\mathrm{w}_{|y}=d\mathrm{u}_y, \:\:d\mathrm{w}_{|z}=d\mathrm{u}_z.
\end{equation}
This form of differentials (which will be confirmed later) is more general than that of differentials (\ref{dif x}, \ref{dif y}). The perpendicular part of the differential (\ref{dif y}) is modified here in a similar way as in the paper \cite{Rebilas 2}.

\section{3. Binary subtraction of 4D and 3D velocities}

Most often, a ordinary operation is a binary operation. Some velocity operations depend on additional parameters, for example operation (\ref {velocity y}) or ternary subtraction, discussed in the next section.
In this article, the terms ``binary'' and ``ternary'' have the nature of proper names borrowed from Oziewicz \cite{binary, Oziewicz} -- they are adequate in the case of 4D, while in 3D they are less precise.

Consider some generalization of the four-vector velocity subtraction (\ref{naiwne}) as a linear combination:
\begin{equation}
\label{4D}
    \omega^{\mu}:=(\upsilon \dsub \nu)^{\mu}:=\lambda_2\:\upsilon^{\mu}-\lambda_1\:\nu^{\mu}.
\end{equation}
In order to determine the $\lambda_1,\lambda_2$ formfactors, we impose two conditions: orthogonality to the subtrahend four-velocity and reduction of the result to ordinary velocity for zero subtrahend three-velocity ($\mathrm{u}_i\equiv\mathrm{u}^i$):
\begin{equation}
\label{warunki}
    \omega\circ \nu:=0,\:\:\:\:\omega^i(\mathrm{v}=0):=\mathrm{u}^i=\upsilon^i/\gamma_{\mathrm{u}}.
\end{equation}
These conditions with the signature $+ - - - (\upsilon_i=-\upsilon^i)$ lead to the value $\lambda_1=1,\lambda_2=c^2/(\upsilon\circ\nu)$ and the final form of the relative binary four-velocity (see \cite{Koczan 2019, groupoid, binary, Bolos}): 
\begin{equation}
\label{binary}
    \omega^{\mu}=(\upsilon \dsub \nu)^{\mu}=\frac{c^2}{\upsilon\circ\nu}\upsilon^{\mu}-\nu^{\mu}.
\end{equation}

STATEMENT 2

The norm of the binary subtraction of the four-velocities (\ref{binary}) with the accuracy of the sign of the signature is equal to the norm of Einstein subtraction (\ref{velocity Einstein}):
\begin{equation}
\label{norma}
||\upsilon \dsub \nu||^2=-|\mathbf{u}\ominus_0\mathbf{v}|^2.
\end{equation}

Proof. The square of the binary four-vector is:
\begin{equation}
\label{kwadrat}
    \omega^{\mu}\omega_{\mu}=\frac{c^6}{(\upsilon\circ\nu)^2}-c^2.
\end{equation}
The easiest way to calculate the square of the subtraction of velocity is in a properly directed coordinate system $xyz$:
\begin{equation}
\begin{split}
\mathrm{u}_x'^2+\mathrm{u}_y'^2=\frac{(\mathrm{u}_x-\mathrm{v})^2+\mathrm{u}_y^2(1-\mathrm{v}^2/c^2)}{(1-\mathrm{u}_x\mathrm{v}/c^2)^2}\\ =\frac{c^2(1-\mathrm{u}_x\mathrm{v}/c^2)^2-(c^2-\mathrm{u}^2)(1-\mathrm{v}^2/c^2)}{(1-\mathrm{u}_x\mathrm{v}/c^2)^2},
\end{split}
\end{equation}
which with the opposite sign equals (\ref{kwadrat}), Q.E.D.

Similarly to the dependence of the four-velocity on the three-velocity (\ref{fourvectors}) or more exact the four-force on the three-force, we can introduce a three-dimensional equivalent for the binary four-velocity:
\begin{equation}
    \omega^{\mu}=(\gamma_{\mathrm{v}} \mathbf{wv}/c, \gamma_{\mathrm{v}} \mathbf{w}).
\end{equation}
The three-dimensional relative binary velocity \cite{Koczan 2020} (jet velocity \cite{Koczan 2019}) can be expressed as follows:
\begin{equation}
\label{jet}
\mathbf{w}=\frac{\mathbf{u}-\mathbf{v}+(\mathbf{u}\times\mathbf{v})\times\mathbf{v}/c^{2}}{1-\mathbf{u}\mathbf{v}/c^{2}}=:\mathbf{u}\ominus_{\perp}\mathbf{v}.
\end{equation}
At the same time, a new three-velocity subtraction operation was introduced that is simpler than operations $\ominus_0$ and $\ominus_{\parallel}$.

STATEMENT 3

Binary 3D subtraction of velocity vectors (\ref{jet}) is of the form:
\begin{equation}
\label{jet factor}
\mathbf{u}\ominus_{\perp}\mathbf{v}=\mathbf{w}=\psi\mathbf{u}-\mathbf{v},
\end{equation}
where the function $\psi$ results from the parallel subtraction according to (\ref{axial x}).

Proof. Using (\ref{axial x}) with (\ref{jet factor}) allows to calculate:
\begin{equation}
\label{psi}
\psi=\frac{1-\mathrm{v}^2/c^2}{1-\mathrm{u}_x\mathrm{v}/c^2}=\frac{1-\mathrm{v}^2/c^2}{1-\mathbf{u}\mathbf{v}/c^2},
\end{equation}
which inserted into (\ref{jet factor}) after simplifications leads to 3D binary subtraction (\ref{jet}), Q.E.D.

The above statement shows that 3D vector subtraction differs from Einstein subtraction in orthogonal component:
\begin{equation}
\label{jet y}
\mathrm{w}_y=(\mathbf{u}\ominus_{\perp}\mathbf{v})_y=\mathrm{u}_y\frac{1-\mathrm{v}^2/c^2}{1-\mathrm{u}_x\mathrm{v}/c^2}.
\end{equation}
Seemingly, the difference relative to (\ref{velocity y}) is not large and does not seem to solve the non-intuitive problem (apart from time dilation) of scaling the perpendicular component.
However, the problem of this scaling disappears if the subtracted velocities have equal components
 $\mathrm{u}_x=\mathrm{v}\equiv\mathrm{v}_x$.

The above property is important in the computation of the 3D binary velocity differential (differential of jet velocity):
\begin{equation}
\label{dif jet x}
d\mathrm{w}_x:=\frac{\partial \mathrm{w}_x}{\partial \mathrm{u}_x}\Big|_{\mathrm{u}_x=\mathrm{v}}d\mathrm{u}_x=\gamma^2_{\mathrm{v}}d\mathrm{u}_x,
\end{equation}
\begin{equation}
\label{dif jet y}
d\mathrm{w}_y:=\frac{\partial \mathrm{w}_y}{\partial \mathrm{u}_x}\Big|_{\mathrm{u}_y=0}d\mathrm{u}_x+\frac{\partial \mathrm{w}_y}{\partial \mathrm{u}_y}\Big|_{\mathrm{u}_x=\mathrm{v}}d\mathrm{u}_y=d\mathrm{u}_y.
\end{equation}
Relative 3D binary velocity differentials coincide with axial velocity differentials as opposed to Einstein subtraction differentials.

\section{4. Ternary subtraction of 4D and 3D velocities}

Based on (\ref{4D},\ref{binary}), we already know that naive subtraction of the four-velocities (\ref{naiwne}) has little specific sense.
However, we can consider such a subtraction projected onto a 3D hypersurface orthogonal to a certain reference four-velocity $\sigma^{\mu}$:  
\begin{equation}
\label{ternary 4D}
    \xi^{\mu}_{(\sigma)}:=(\upsilon \triangleminus_{\sigma} \nu)^{\mu}:=\lambda\:P_{\perp\beta}^{\mu}(\sigma)(\upsilon^{\beta}-\nu^{\beta}).
\end{equation}
It is therefore a ternary velocity subtraction dependent on the three four-velocities. The orthogonal projection operator has the form $P_{\perp\beta}^{\mu}=\delta^{\mu}_{\beta}-\sigma^{\mu}\sigma_{\beta}/c^2$. In order to calculate the form-factor $\lambda$ for the sake of simplicity, we will limit ourselves to the three-dimensional approach, assuming that $\sigma^{\mu}=(c^2,0,0,0)$. This leads to a 3D relative ternary velocity:
\begin{equation}
\label{binary 3D}
    \mathbf{W}:=\mathbf{u} \ominus_{\wedge} \mathbf{v}:=\lambda  \ (\gamma_{\mathrm{u}}\mathbf{u}-\gamma_{\mathrm{v}}\mathbf{v}).
\end{equation}
The ternarity of this operation is implicitly hidden in choice $\sigma$. Regardless of the choice of $\lambda$, this operation is not equal to any of the previously discussed. Nevertheless, in the case of axial (parallel vectors) a condition can be imposed on $\lambda$ conforming to standard subtraction:
\begin{equation}
\label{warunek lambda}
    \mathrm{u} \ominus_{\wedge} \mathrm{v}:=\mathrm{u} \ominus \mathrm{v}\:\:\:\:(\mathbf{u}\parallel\mathbf{v}).
\end{equation}
Unfortunately, the condition cannot be extended to each case for a parallel component. The given definition condition allows, however, to calculate:
\begin{equation}
\label{lambda}
\begin{split}
\lambda=\frac{\mathrm{u}-\mathrm{v}}{(1-\frac{\mathrm{uv}}{c^2})(\gamma_{\mathrm{u}}\mathrm{u}-\gamma_{\mathrm{v}}\mathrm{v})}
=\frac{(\mathrm{u}-\mathrm{v})(\gamma_{\mathrm{u}}^{-1}+\gamma_{\mathrm{v}}^{-1})}{(1-\frac{\mathrm{uv}}{c^2})(\mathrm{u}-\mathrm{v}+\frac{\gamma_{\mathrm{u}}}{\gamma_{\mathrm{v}}}\mathrm{u}-\frac{\gamma_{\mathrm{v}}}{\gamma_{\mathrm{u}}}\mathrm{v})}=\\
=\frac{\gamma_{\mathrm{u}}^{-1}+\gamma_{\mathrm{v}}^{-1}}{(1-\frac{\mathrm{uv}}{c^2})(1+\gamma_{\mathrm{u}}\gamma_{\mathrm{v}}(1+\frac{\mathrm{uv}}{c^2}))}=
\frac{\gamma_{\mathrm{u}}^{-1}+\gamma_{\mathrm{v}}^{-1}}{1-\frac{\mathrm{uv}}{c^2}+\gamma_{\mathrm{u}}\gamma_{\mathrm{v}}(1-\frac{\mathrm{u^2v^2}}{c^2})}.
\end{split}
\end{equation}
We will now make a generalization of this expression for the case of any velocities, assuming that the square of the velocity concerns the square of the norm ($\mathrm{u}^2=\mathbf{u}^2$) and the expression of the first degree in a given velocity denotes its component ($\mathrm{u}=\mathrm{u}_x, \:\mathrm{uv}=\mathrm{u}_x\mathrm{v}=\mathbf{uv}$). The validity of these assumptions will be further confirmed in the form of the Lemma proof. Using that  $\mathrm{u}^2/c^2=1-\gamma_{\mathrm{u}}^{-2}$ we get:
\begin{equation}
\label{lambda 2}
\lambda
=\frac{\gamma_{\mathrm{u}}^{-1}+\gamma_{\mathrm{v}}^{-1}}{1-\frac{\mathbf{uv}}{c^2}+\gamma_{\mathrm{u}}\gamma_{\mathrm{v}}(1-\frac{\mathbf{u}^2\mathbf{v}^2}{c^4})}
=\frac{\gamma_{\mathrm{u}}+\gamma_{\mathrm{v}}}{\gamma_{\mathrm{u}}\gamma_{\mathrm{v}}(1-\frac{\mathbf{uv}}{c^2})+\gamma_{\mathrm{u}}^2+\gamma_{\mathrm{v}}^2-1}.
\end{equation}
The covariant equivalent of this expression is:
\begin{equation}
\label{lambda 3}
\lambda
=\frac{c^2\sigma\circ\upsilon+c^2\sigma\circ\nu}{c^2\upsilon\circ\nu+(\sigma\circ\upsilon)^2+(\sigma\circ\nu)^2-c^4}.
\end{equation}
Finally, the relative ternary four-velocity becomes:
\begin{equation}
\label{ternary full 4D}
\xi^{\mu}_{(\sigma)}=(\upsilon \triangleminus_{\sigma} \nu)^{\mu}=\frac{\sigma\circ(\upsilon+ \nu)(c^2\delta^{\mu}_{\beta}-\sigma^{\mu}\sigma_{\beta})(\upsilon^{\beta}-\nu^{\beta})}{(\sigma\circ\upsilon)^2+(\sigma\circ\nu)^2+c^2\upsilon\circ\nu -c^4}.
\end{equation}
This velocity, written differently and derived differently, was first published by Oziewicz in 2004 \cite{binary, Oziewicz, Koczan 2019}. Ternary subtraction is a generalization of binary subtraction which is a special case ($\sigma=\nu$):
\begin{equation}
 \upsilon \dsub \nu=\upsilon \triangleminus_{\nu} \nu.
\end{equation}
 
Another special case ($\sigma_i=0$) is the already introduced three-dimensional ternary relative velocity:
\begin{equation}
\label{ternary full 3D}
\mathbf{W}=\mathbf{u}\ominus_{\wedge}\mathbf{v}=\frac{(\gamma_{\mathrm{u}}+\gamma_{\mathrm{v}})(\gamma_{\mathrm{u}}\mathbf{u}-\gamma_{\mathrm{v}}\mathbf{v})}
{\gamma_{\mathrm{u}}\gamma_{\mathrm{v}}(1-\frac{\mathbf{uv}}{c^2})+\gamma_{\mathrm{u}}^2+\gamma_{\mathrm{v}}^2-1}.
\end{equation}
This velocity, written and derived somewhat differently, first appeared in 2012 in Dragan \cite{Dragan} lectures notes, but has not been published so far either in the form of an article or in the form of a preprint (except for the Polish lectures). The poster \cite{Koczan 2020} provides a visualized interpretation of the 3D ternary velocity using the so-called laufer/bishop method.

Both 4D and 3D ternary subtractions are antisymmetric:
\begin{equation}
\upsilon \triangleminus_{\sigma} \nu=-\nu \triangleminus_{\sigma} \upsilon\ \ , \ \ \mathbf{u}\ominus_{\wedge} \mathbf{v}=-\mathbf{v}\ominus_{\wedge} \mathbf{u}.
\end{equation}
Other than the one-dimensional velocity subtractions considered in this paper do not have this property - including the Einstein subtraction. The differences do not end there:

STATEMENT 4

Ternary 3D subtraction, with the exception of parallel velocities, does not generally coincide with Einstein subtraction in the subtrahend direction, nor do the norms of the results of these operations:
\begin{equation}
 (\mathbf{u}\ominus_{\wedge} \mathbf{v})_{\mathbf{v}} \not\equiv (\mathbf{u}\ominus_0 \mathbf{v})_{\mathbf{v}},\ \ \ \ |\mathbf{u}\ominus_{\wedge} \mathbf{v}| \not\equiv |\mathbf{u} \ominus_0 \mathbf{v}|.
\end{equation}

Proof. It is enough to indicate one example for which the considered equations do not exist. So let us consider perpendicular velocities with equal values: $\mathrm{u}=\mathrm{v}=\mathrm{v}_x=\mathrm{u}_y=0.6c$. Einstein's subtracting leads to the components of the velocity vector:
\begin{equation}
\mathrm{u}'_x=\frac{0-0.6c}{1-0\times0.6}=-0.6c,
\end{equation}
\begin{equation}
\mathrm{u}'_y=0.6c\frac{\sqrt{1-0.6^2}}{1-0\times0.6}=0.48c,
\end{equation}
which value is $\mathrm{u}'\approx0.7684c$. Whereas the components of the 3D ternary velocity are:
\begin{equation}
\mathrm{W}_x=\frac{(\frac{5}{4}+\frac{5}{4})\times(\frac{5}{4}\times0 -\frac{5}{4}\times\frac{3}{5}c)}{\frac{5}{4}\times\frac{5}{4}\times(1-0)+\frac{5^2}{4^2}+\frac{5^2}{4^2}-1}=-\frac{30}{59}c\approx-0.5085c,
\end{equation}
\begin{equation}
\mathrm{W}_y=\frac{(\frac{5}{4}+\frac{5}{4})\times(\frac{5}{4}\times\frac{3}{5}c -\frac{5}{4}\times0)}{\frac{5}{4}\times\frac{5}{4}\times(1-0)+\frac{5^2}{4^2}+\frac{5^2}{4^2}-1}=\frac{30}{59}c\approx0.5085c,
\end{equation}
which gives it value $\mathrm{W}\approx0.7191c$. Thus, neither the $x$ components nor the values of the considered velocities are equal, Q.E.D.  

Statement 4 would suggest that the 3D ternary velocity is weakly anchored in the Lorentz transformation. However, nothing could be more wrong, as shown below (see \cite{Dragan}):

LEMMA

The Lorentz boost (Einstein subtraction) with a ternary 3D velocity  for a velocity, which is minuend ternary subtraction, restores ternary subtrahend velocity:
\begin{equation}
\label{lemat}
 \mathbf{u}\ominus_0 \mathbf{W}=\mathbf{u}\ominus_0 (\mathbf{u}\ominus_{\wedge} \mathbf{v})=\mathbf{v}.
\end{equation}
In other words, in the above sense, 3D ternary subtraction is a left-inverse operation of Einstein subtraction.

Proof. Proving the Lemma thesis is computationally complicated and it cannot be done just by boost in the $x$ direction. A significant difficulty is calculating the Lorentz factor for velocity $\mathbf{W}$: 
\begin{equation}
\frac{1}{\gamma^2_{\mathrm{W}}}=1-\frac{\mathrm{W}^2}{c^2}
=\frac{\big(\gamma_{\mathrm{u}}\gamma_{\mathrm{v}}\frac{\mathbf{uv}}{c^2}+\gamma_{\mathrm{u}}\gamma_{\mathrm{v}}+1\big)^2}{\big(\gamma_{\mathrm{u}}\gamma_{\mathrm{v}}(1-\frac{\mathbf{uv}}{c^2})+\gamma_{\mathrm{u}}^2+\gamma_{\mathrm{v}}^2-1\big)^2},
\end{equation}
which can be written more compactly:
\begin{equation}
\frac{1}{\gamma_{\mathrm{W}}}+1=(\gamma_{\mathrm{u}}+\gamma_{\mathrm{v}})\lambda.
\end{equation}
The main calculations can now be made:
\begin{equation}
\label{liczenie v}
\mathbf{u}\ominus_0\mathbf{W}
=\frac{\frac{1}{\gamma_{\mathrm{W}}}\mathbf{u}-\mathbf{W}+\frac{\mathbf{u}\mathbf{W}/c^2}{1+\gamma^{-1}_{\mathrm{W}}}\mathbf{W}}{1-\frac{\mathbf{uW}}{c^2}}
=\frac{B\:\mathbf{u}+C\:\mathbf{v}}{1-\frac{\mathbf{uW}}{c^2}}.
\end{equation}
Now it is enough to calculate the coefficients $B$ and $C$:
\begin{equation}
B=(\gamma_{\mathrm{u}}+\gamma_{\mathrm{v}})\lambda-1-\gamma_{\mathrm{u}}\lambda+\frac{\gamma^2_{\mathrm{u}}-1-\gamma_{\mathrm{u}}\gamma_{\mathrm{v}}\frac{\mathbf{uv}}{c^2}}{\gamma_{\mathrm{u}}+\gamma_{\mathrm{v}}}\lambda=0,
\end{equation}
\begin{equation}
C=\frac{\gamma_{\mathrm{v}}(\gamma_{\mathrm{u}}\gamma_{\mathrm{v}}+1+\gamma_{\mathrm{u}}\gamma_{\mathrm{v}}\frac{\mathbf{uv}}{c^2})}{\gamma_{\mathrm{u}}(\gamma_{\mathrm{u}}+\gamma_{\mathrm{v}})}\lambda=1-\frac{\mathbf{uW}}{c^2}.
\end{equation}
Based on these calculations, the value of the expression (\ref{liczenie v}) is $\mathbf{v}$, Q.E.D.

The ordinary understood inverse Lorentz transformation is the right-hand inverse (according to the adopted notation convention):
\begin{equation}
 \mathbf{u}\ominus_0\mathbf{v}=\mathbf{u}' \ \ \rightarrow \ \ 
 \mathbf{u}=\mathbf{u}'\ominus_0(-\mathbf{v}).
\end{equation}
On the other hand, the ternary operation can be understood as a left-inverse operation:
\begin{equation}
 \mathbf{u}\ominus_0 \mathbf{W}=\mathbf{v} \ \ \rightarrow \ \ \mathbf{W}=\mathbf{u}\ominus_{\wedge} \mathbf{v}.
\end{equation}
Like Einstein subtraction, left-hand application is the left-hand inverse of ternary action (\ref{lemat}). It is worth writing it explicitly in both order:
\begin{equation}
 \mathbf{u}\ominus_0(\mathbf{u}\ominus_{\wedge}\mathbf{v})=\mathbf{u}\ominus_{\wedge}(\mathbf{u}\ominus_0\mathbf{v})=\mathbf{v}.
\end{equation}

To sum up, the main reason for the difference between the 3D ternary velocity subtraction from the Einstein subtraction is the noncommutativity of velocity  composition (and Lorentz group). The impact of the lack of associative velocity  composition is not directly apparent here, and the Lorentz group is associative by definition.

The differential of the 3D ternary velocity, which is crucial for this paper, remains to be calculated. The ternary velocity is so homogeneous due to its components that we can immediately calculate the vector differential:
\begin{equation}
\label{dW}
d\mathbf{W}
:=\frac{\partial\mathbf{W}}{\partial\mathrm{u}_i}\Big|_{\mathbf{u}=\mathbf{v}}d\mathrm{u}_i
=d\mathbf{u}+\frac{\gamma^2_{\mathrm{v}}}{c^2}\mathbf{v}(\mathbf{v}d\mathbf{u})=d\mathbf{u}_{\perp}+\gamma^2_{\mathrm{v}} d\mathbf{u}_{\parallel}.
\end{equation}
As before, the velocity differential $\mathbf{v}$ would only differ in sign and names of symbols. Thus, as can be seen, the 3D ternary velocity differential is consistent with the 3D axial velocity differential and the 3D binary velocity differential. This compatibility is not accidental and reflects a class of differentially equivalent operations to which Einstein's subtraction does not belong.

\section{5. Relativistic 4D and 3D acceleration}

We will start with the 4D approach as it will pave the way to the main original 3D result. The standard four-acceleration is derivative of the four-velocity respect to self-time \cite{Wroblewski}:
\begin{equation}
a^{\mu}=\frac{d\upsilon^{\mu}}{d\tau}=(a^0,\ \vec{a})=\big(\gamma^4_{\mathrm{u}}\mathbf{au}/c,\ \gamma^2_{\mathrm{u}}\mathbf{a}+\gamma^4_{\mathrm{u}}(\mathbf{au})\mathbf{u}/c^2\big).
\end{equation}
Despite the simple definition, four-acceleration depends on the ordinary three-acceleration $\mathbf{a}$ in a very complicated way. The same result can be obtained for a binary four-velocity if we identify the subtracted four-velocities after derivative calculated:
\begin{equation}
a^{\mu}=\frac{d\omega^{\mu}}{d\tau}(\tau_0, \tau)|_{\tau=\tau_0}
=\frac{d\big(\upsilon(\tau)\dsub\upsilon(\tau_0)\big)^{\mu}}{d\tau}\Big|_{\tau=\tau_0}.
\end{equation}
This is equivalent to writing the derivative definition, in which the ordinary subtraction is replaced by a binary 4D subtraction:
\begin{equation}
a^{\mu}
=\lim_{\Delta\tau \rightarrow 0}\frac{\big(\upsilon(\tau_0+\Delta\tau)\dsub\upsilon(\tau_0)\big)^{\mu}}{\Delta\tau}.
\end{equation}
Thanks to this notation, it is clear that when calculating the acceleration (or differential) from the relative velocity, the subtracted velocities $\nu^{\mu}$ and $\upsilon^{\mu}$ should tending to each other.

For the ternary four-velocity, there is also the third reference four-velocity $\sigma^{\mu}$. If this four-velocity were to take the same value as the previous two, then the four-acceleration calculations coincided with the binary calculations. Thus, it is worth treating $\sigma^{\mu}$ as the four-velocity of the independent selected frame of reference,  in which the moving time will be measured and the acceleration calculated. Time in the inertial frame $\sigma$ relative to selftime of body with a four-velocity  $\upsilon^{\mu}(\tau)$ is described by time dilation $dT=(\sigma\circ\upsilon/c^2) d\tau=d\tau/\lambda(\sigma,\upsilon,\upsilon)$. This leads to the definition of the four-acceleration from 4D ternary subtraction:
\begin{equation}
\begin{split}
\mathcal{A}^{\mu}_{(\sigma)}
:=\frac{d\xi^{\mu}_{(\sigma)}}{dT}
:=\lim_{\Delta\tau \rightarrow 0}\frac{\big(\upsilon(\tau_0+\Delta\tau)\triangleminus_{\sigma}\upsilon(\tau_0)\big)^{\mu}}{\Delta\tau/\lambda}=\\
=\frac{c^4}{(\sigma\circ\upsilon)^2}\big(a^{\mu}-(\sigma\circ a)\sigma^{\mu}/c^2\big).
\end{split}
\end{equation}
The ternary four-acceleration is thus a properly normalized projection of the ordinary four-acceleration. For $\sigma=\upsilon$ four-acceleration $\mathcal{A}^{\mu}_{(\sigma)}$ coincides with four-acceleration $a^{\mu}$, and for $\sigma_i=0$ it amounts to a three-dimensional relativistic acceleration $\mathbf{A}$ -- in the same way that the 4D ternary velocity comes down to the 3D ternary velocity.

It's time to define the 3D relativistic acceleration. For definition, we need a velocity differential based on the difference of velocities, that is on relative velocity. Apart from the ordinary subtraction, we have Einstein subtraction and three other methods of 3D subtraction: axial, binary and ternary. Einstein subtraction leads to a velocity differential in the system of instantaneous rest of the body, which only allows the determination of the rest acceleration $\mathbf{a}_0=d\mathbf{u}'/d\tau$ \cite{Koczan 2019}. The rest acceleration is not an acceleration {\it per se}, so they remain
consistent differentials of axial, 3D binary and ternary  velocities. This velocity differential will be called the relativistic differential (of velocity) and will be denoted as follows:
\begin{equation}
\label{rozniczka D}
D\mathbf{u}
:=d\mathbf{w}_|=d\mathbf{w}=d\mathbf{W}
=d\mathbf{u}_{\perp}+\gamma^2_{\mathrm{u}} d\mathbf{u}_{\parallel}.
\end{equation}

STATEMENT 5

The velocity relativistic differentiation operation is a differentiation operation multiplied on both sides by the gamma factor and its inverse (on the left):
\begin{equation}
D=\gamma^{-1}\times d \times \gamma.
\end{equation}
Proof:
\begin{equation}
D\mathbf{u}=\gamma^{-1}_{\mathrm{u}}d(\gamma_{\mathrm{u}}\mathbf{u})
=d\mathbf{u}+\gamma^2_{\mathrm{u}}\mathbf{u}(\mathbf{u}d\mathbf{u})/c^2,
\end{equation}
which is equal to (\ref{rozniczka D}) by virtue of (\ref{dW}), Q.E.D.

\begin{figure}[h!]
\centering
\includegraphics[width=8cm]{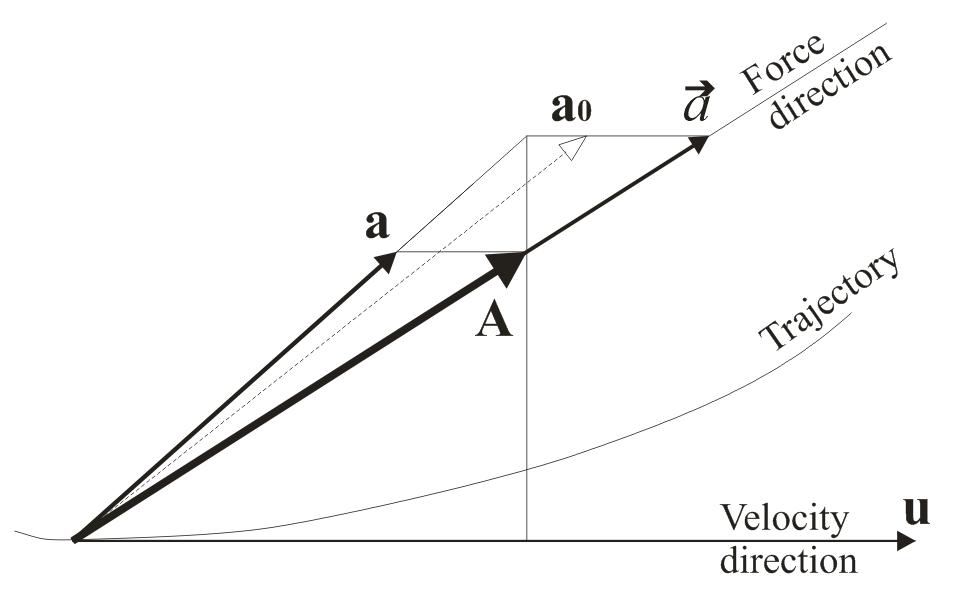}
\caption{Diagram showing the relations of the main 3D acceleration vectors. The new  relativistic acceleration $\mathbf{A}$ is a parallel projection of ordinary acceleration $\mathbf{a}$ along the velocity on the direction of force (or spatial part $\vec{a}$ of the four-acceleration $a^{\mu}$). Thanks to this, the construction of acceleration $\mathbf{A}$ is simpler than the construction of rest acceleration $\mathbf{a}_0$}
\label{diagram}
\end{figure}

The key relativistic 3D acceleration for this paper can now be defined:
\begin{equation}
\label{A}
\mathbf{A}:=\frac{D\mathbf{u}}{dt}:=\lim_{\Delta t\rightarrow 0}\frac{\mathbf{u}(t+\Delta t)\ominus_{\wedge} \mathbf{u}(t)}{\Delta t},
\end{equation}
where antisymmetric ternary subtraction can be replaced by binary or axial subtraction. This definition leads to the formula (see Fig.\ref{diagram}):
\begin{equation}
\mathbf{A}
=\mathbf{a}+\gamma^2_{\mathrm{u}}\mathbf{u}(\mathbf{u}\mathbf{a})/c^2
=\mathbf{a}_{\perp}+\gamma^2_{\mathrm{u}}\mathbf{a}_{\parallel},
\end{equation}
where $\mathbf{a}=d\mathbf{u}/dt$ is the ordinary acceleration.

Four- acceleration can now be expressed simply by acceleration $\mathbf{A}$:
\begin{equation}
a^{\mu}=(a^0,\ \vec{a})=(\gamma^2_{\mathrm{u}}\mathbf{Au}/c,\ \gamma^2_{\mathrm{u}}\mathbf{A}).
\end{equation}
Thanks to this, also the ternary four-acceleration $\mathcal{A}^{\mu}_{(\sigma)}$ can be expressed as acceleration $\mathbf{A}$. However, this is a more complex expression that simplifies for $\sigma_i=0$:
\begin{equation}
\mathcal{A}^{\mu}_{(\sigma_i=0)}=(0,\mathbf{A}).
\end{equation}

\section{6. Conclusion}
It has been shown and proven in the paper that within the Lorentz group it is possible to reasonably subtract velocities in a different way than it was established in SR. First of all, it can be done in an antisymmetric way or in a four-vector way. For example, the antisymmetric operation is the left-hand reciprocal of the Lorentz boost. Velocity subtraction operations made it possible to define different 4D and 3D relative velocities. Due to the differential equivalence of these subtraction operations, unambiguous and original 3D relativistic acceleration was introduced. This relativistic acceleration is important for simplifying the dynamics of the SR \cite{Koczan 2019, Koczan 2020}. Whereas in this article, the relativistic acceleration has been generalized to the 4D ternary four-acceleration.

\bibliographystyle{unsrt}

\begin{thebibliography}{99}

\bibitem{Ahmad} M. Ahmad, ``Ambiguity in Lorentz Transformation and 
Reciprocal Symmetric Transformation as the Answer'',  {\it International Journal of Reciprocal Symmetry and Theoretical Physics} \textbf{1}(2), pp. 69--79 (2014), DOI: 10.15590/ijrstp.

\bibitem{Ahmad Alam} M. Ahmad, M. S. Alam, ``Non-Associativity of Lorentz
Transformation and Associative Reciprocal Symmetric Transformation'',  {\it International Journal of Reciprocal Symmetry and Theoretical Physics} \textbf{1}(1), pp. 9--19 (2014), DOI: 10.15590/ijrstp/2014/v1i1/53720.

\bibitem{Barrett} J.F. Barrett, ``Review of problems of dynamics in the hyperbolic theory of Special Relativity'', PIRT Conf., Imperial Coll., London, Proceedings ISBN 1 873 694 07 5, PD Publications (Liverpool), pp. 17--30 (2002). 

\bibitem{Bolos} V.J. Bol\'os, ``Intrinsic definitions of ``relative velocity'' in
General Relativity'', arXiv:0506032v1 (2005), {\it Communications in Mathematical Physics} \textbf{273}, pp. 217--236 (2007),
DOI: 10.1007/s00220-007-0248-9.


\bibitem{Cannoni} M. Cannoni, ``Lorentz invariant relative velocity and relativistic binary collisions'', arXiv:1605.00569v2 (2016).

\bibitem{Dragan} A. Dragan, ``Niezwykle szczeg\'olna teoria wzgl\c{e}dno\'sci. Roz. 3. Obr\'ot Thomasa--Wignera  (Unusually special theory of relativity. Chap. 3. Thomas--Wigner rotation)'', monograph -- lecture notes, www.researchgate.net/publication/265887295 (2012).

\bibitem{Dragan AJP} A. Dragan, T. Odrzyg\'o\'zd\'z, ``Half-page derivation of the Thomas precession'', {\it American Journal of Physics} \textbf{81} (8), 631 (2013), DOI: 10.1119/1.4807564.

\bibitem{Einstein} A. Einstein, ``Zur Elektrodynamik bewegter K\"orper (On the Electrodynamics of Moving Bodies)'',
{\it Annalen der Physik} \textbf{17}, pp. 891--921, June 30 (1905).

\bibitem{Mexico} M. Fern\'andez-Guasti, ``Alternative realization for the composition of relativistic
velocities'', Proc. of SPIE Vol. 8121, 812108 (2011), DOI: 10.1117/12.894342.

\bibitem{Fock} V. Fock, {\it The Theory of Space, Time and  Gravitation}, 1st ed. 1959, 2nd rev. ed., translated by N. Kemmer, Pergamon Press (1964).

\bibitem{Kocik} J. Kocik, ``Making sense of relativistic composition of velocities'', arXiv:1910.06785v1 (2019).

\bibitem{Koczan 2019} Koczan, G.M.: ``New definitions of 3D acceleration and inertial mass not violating F={\it M}A in the Special Relativity'', arXiv:1909.09084v3 (v1: 2019), {\it Results in Physics} \textbf{24}, 104121 (2021), DOI: 10.1016/j.rinp.2021.104121.

\bibitem{Koczan 2020} Koczan, G.M.: ``The new definition of three-dimensional relativistic acceleration and its consequences within the SR'', https://100lat.ptf.net.pl/plakaty/0286-plakat\_en-1be04124.pdf,         researchgate.net/publication/344670531 (DOI: 10.13140/RG.2.2.27955.45603),   poster for XLVI Extraordinary Congress of Polish Physicists on the 100 Years of the Polish Physical Society, Warsaw, October 16-18 (2020).

\bibitem{Minkowski} H. Minkowski, ``Die Grundgleichungen für die elektromagnetischen Vorg\"ange in bewegten K\"orpern (The Fundamental Equations for Electromagnetic Processes in Moving Bodies)'', {\it Nachrichten von der Gesellschaft der Wissenschaften zu G\"ottingen, Mathematisch-Physikalische Klasse}, pp. 53--111, Berlin (1908). 

\bibitem{binary} Z. Oziewicz, ``How do you add relative velocities?'', {\it Group Theoretical Methods in Physics, Conference Series} \textbf{185}, pp. 439--444,  CRC Press (2004), DOI: 10.1201/9781482269185 .

\bibitem{Oziewicz} Z. Oziewicz, ``Ternary relative velocity'', arXiv:1104.0682v1 (2011).

\bibitem{groupoid} Z. Oziewicz, ``Relativity groupoid, instead of relativity group'', {\it International Journal of Geometric Methods in Modern Physics} \textbf{04} (05), pp. 739--749 (2007), DOI: 10.1142/S0219887807002260.

\bibitem{Rebilas} K. R\c{e}bilas, ``Derivation of the relativistic
momentum and relativistic equation of motion from Newton's second law and Minkowskian space-time
geometry'', {\it Apeiron} \textbf{15} (3), July (2008).

\bibitem{Rebilas 2} K. R\c{e}bilas, ``Comment on ‘Elementary analysis of
the special relativistic combination of velocities, Wigner rotation and Thomas precession’ '', {\it European Journal of Physics} \textbf{34}, pp. L55--L61 (2013), DOI: 10.1088/0143-0807/34/3/L55.

\bibitem{Rebilas 3} K. R\c{e}bilas, ``Lorentz-invariant three-vectors and alternative formulation of relativistic dynamics'', {\it American Journal of Physics} \textbf{78}, 294 (2010), DOI: 10.1119/1.3258203.

\bibitem{Rindler} W. Rindler, {\it Relativity: Special, General and Cosmological}, Oxford University Press (2007), DOI: 10.1007/s10714-007-0401-y.

\bibitem{Ungar} A. A. Ungar, ``The Relativistic Noncommutative Nonassociative Group of Velocities and the Thomas Rotation'', {\it Results in Mathematics} \textbf{16}, pp. 168--179 (1989), DOI: 10.1007/BF03322653.

\bibitem{Wroblewski} A. K. Wr\'oblewski, J. A. Zakrzewski, {\it Wst\c{e}p do fizyki (Introduction to Physics)} Vol. 1, PWN (1976, 1984).


\end{thebibliography}

\end{document}